\documentclass{jpp}
\usepackage{graphicx}

\newcommand{\citep}[1]{[\cite{#1}]}

\title{Lagrangian Statistics of Navier-Stokes- and MHD-Turbulence}

\author{H.Homann and R. Grauer $^1$ \\
        A. Busse and W.C. M\"uller $^2$}

\affiliation{
  $^1$Theoretische Physik I, Ruhr-Universit\"at Bochum, Germany \\[\affilskip]
  $^2$MPI for Plasma Physics, Garching, Germany}

\date{23 November 2006}

\begin{document}

\maketitle

\begin{abstract}
  We report on a comparison of high-resolution numerical simulations
  of Lagrangian particles advected by incompressible turbulent hydro-
  and magnetohydrodynamic (MHD) flows.  Numerical simulations were
  performed with up to $1024^3$ collocation points and 10 million
  particles in the Navier-Stokes case and $512^3$ collocation points
  and 1 million particles in the MHD case. In the hydrodynamics case
  our findings compare with recent experiments from Mordant et
  al.~\citep{mordant-leveque-etal:2004} and Xu et
  al.~\citep{xu-bourgain-etal:2005}. They differ from the simulations
  of Biferale et al.~\citep{biferale-etal:2004} due to differences of
  the ranges choosen for evaluating the structure functions. In
  Navier-Stokes turbulence intermittency is stronger than predicted by
  a multifractal approach of \citep{biferale-etal:2004} whereas in MHD
  turbulence the predictions from the multifractal approach are more
  intermittent than observed in our simulations.  In addition, our
  simulations reveal that Lagrangian Navier-Stokes turbulence is more
  intermittent than MHD turbulence, whereas the situation is reversed
  in the Eulerian case. Those findings can not consistently be
  described by the multifractal modeling. The crucial point is that
  the geometry of the dissipative structures have different
  implications for Lagrangian and Eulerian intermittency.  Application
  of the multifractal approach for the modeling of the acceleration
  PDFs works well for the Navier-Stokes case but in the MHD case just
  the tails are well described.
\end{abstract}



\maketitle

\noindent
\section{Introduction} Lagrangian statistics of turbulent flows has
undergone a rapid development in the last 6 years due to enormous
progress in experimental techniques measuring particle trajectories.
Particle tracking velocimetry has been used for moderate Reynolds
numbers by \citep{ott-mann:2000}. However, the techniques developed in
Cornell
\citep{porta-voth-etal:2000,porta-voth-etal:2001a,voth-porta-etal:2001b}
and Lyon~\citep{mordant-metz-etal:2001,mordant-leveque-etal:2004}
allowed the measurements of probability density functions (PDFs) of
velocity increments which triggered a renewed interest in the
theoretical understanding of Lagrangian statistics. A very promising
approach based on a Markovian closure was recently introduced in
\citep{friedrich:2003}. This approach is not readily applicable to
magnetohydrodynamic (MHD) turbulence, since the distribution function
depends not only on velocity, space and initial condition but in
addition on the Jacobian.  Although work in this direction is in
progress, we compare our simulations to a phenomenological model of
Lagrangian statistics in Navier-Stokes turbulence introduced by
Biferale et al.~\citep{biferale-etal:2004}. Our findings show increased
intermittency in Navier-Stokes flows, such that the structure
functions agree with recent experimental data from the two
experimental
groups~\citep{mordant-leveque-etal:2004,xu-bourgain-etal:2005}.  On the
other hand, our MHD simulations are less intermittent than the
predictions from a multifractal model. \\

\noindent
\section{Numerical Methods} The Lagrangian particle trajectories were
obtained by two slightly different parallel spectral codes (Garching
and Bochum) based on the spectral code used in
\citep{mueller-biskamp:2000}.  The velocity and magnetic field was
evaluated at the particle positions using either trilinear or tricubic
interpolation.  Contrary to the simulations of
\citep{biferale-etal:2004}, we found that tricubic interpolation
captures especially trajectories with high acceleration more precise
than linear interpolation, a conclusion which was drawn already 20
years ago (see \citep{yeung:2002} and the discussion therein). However,
the effect of the slightly different trajectories has only a minor
effect on the tails of the acceleration PDFs with the tendency that
the PDF calculated with tricubic interpolation is slightly more
intermittent than the corresponding one calculated with trilinear
interpolation. The simulations presented here use a tricubic
interpolation for obtaining the velocities at the particle positions. \\

\noindent
\section{Navier-Stokes turbulence} We performed a set of simulations
for the Navier-Stokes-equations with two resolutions, $512^3$ and
$1024^3$ collocation points and 1 million and 10 million particles,
respectively, in order to obtain reliable statistical results
within a few large eddy-turnover times.

Parameters of all simulations (Navier-Stokes and MHD) are summarized
in Table~\ref{table1}.  Here we used the same conventions as described
in \citep{frischbook:1995}.  To get to a stationary turbulent state, we
started with randomly distributed Fourier-modes without driving.
After the turbulence has been fully developed, the low mode number
modes were kept constant. Particles with initially homogeneous random
positions were injected when a stationary state was reached. For the
simulation with $1024^3$ collocation points, we started from the
stationary turbulent state with $512^3$ collocation points. The
relaxation to a new stationary state took about one large-eddy
turnover time.  After this period, the particles were injected.

The choice of the parameters for Run1, Run4 and Run5 and therefore the
resulting Taylor microscale Reynolds number was motivated by the
standard procedure to choose the dissipation length $l_d$ smaller than
the grid spacing~\citep{yeung-pope:1989}.  However, recent
investigations of how dissipative structures like shocks, tubes and
sheets enter the dissipation range
\citep{laessig:2000,yakhot-screenivasan:2005,schumacher:2006} suggest a
more conservative choice which was realized in Run2 and Run3.

\begin{table}
  \centering
  \begin{tabular}{lrrrrr}
                & Run1               & Run2               & Run3               & Run4               & Run5                \\ \hline
    $R_\lambda$ &                 190&                 122&                 178&                 187&                 234 \\
    $u_0$       &                0.82&                0.16&                0.16&                0.48&                0.22 \\
    $\epsilon_k$&              $0.23$&  $2.1\cdot 10^{-3}$&    $2\cdot 10^{-3}$&                 0.1&   $1 \cdot 10^{-2}$ \\
    $\epsilon_m$&                   -&                   -&                   -&                0.15&  $1.5\cdot 10^{-2}$ \\
    $\nu=\eta$  &    $8\cdot 10^{-4}$&  $  3\cdot 10^{-4}$&  $1.5\cdot 10^{-4}$&    $5\cdot 10^{-4}$&  $1.5\cdot 10^{-3}$ \\
    dx          &$12.27\cdot 10^{-3}$&$12.27\cdot 10^{-3}$&$ 6.14\cdot 10^{-3}$&$12.27\cdot 10^{-3}$&$12.27\cdot 10^{-3}$ \\
    $l_d$       &  $6.9\cdot 10^{-3}$&  $1.1\cdot 10^{-2}$&  $6.4\cdot 10^{-3}$&  $5.9\cdot 10^{-3}$&  $4.3\cdot 10^{-3}$ \\
    $\tau_d$    &  $5.9\cdot 10^{-2}$&  $3.7\cdot 10^{-1}$&  $2.8\cdot 10^{-1}$&  $7.1\cdot 10^{-2}$&  $1.2\cdot 10^{-1}$ \\
    $L$         &                 2.4&                 1.9&                   2&                 2.4&                 2.5 \\
    $T_L$       &                 2.9&                  12&                  11&                   5&                 6.3 \\
    $T/T_L$     &                10.3&                   5&                   2&                 4.7&                 1.8 \\
    $N^3$       &            $ 512^3$&            $ 512^3$&            $1024^3$&            $ 512^3$&             $512^3$ \\
    $N_p$       &   $ 1.18\cdot 10^6$&   $ 1   \cdot 10^6$&    $10  \cdot 10^6$&   $ 1.18\cdot 10^6$&       $1\cdot 10^6$
  \end{tabular}
  \caption{Parameters of the numerical simulations. 
    $R_\lambda$: Taylor microscale Reynolds number $\sqrt{15 u_0 L/\nu}$,
    $u_0 = \sqrt{2/3 E_k}$, $E_k$: kinetic energy, $E_m$: magnetic energy, $E = E_k + E_m$,
    $\epsilon_k$: kinetic energy dissipation rate,
    $\epsilon_m$: magnetic energy dissipation rate,
    $\epsilon = \epsilon_k + \epsilon_m$,
    $\nu$: viscosity, 
    $\eta$: resistivity,
    $l_d$: dissipation lengthscale $(\nu^3/\epsilon_k)^{1/4}$,
    $\tau_d$: Kolmogorov time scale $(\nu/\epsilon_k)^{1/2}$,
    $L = (2/3 E)^{3/2}/\epsilon$: integral scale,
    $T_L = L/u_0$: large-eddy turnover time,
    $T$: total integration time,
    $N^3$: number of collocation points,
    $N_p$: number of particles, Navier-Stokes simulations: Run1-Run3, MHD simulations: Run4, Run5}
\label{table1}
\end{table}

\begin{figure}
  \centering
  \includegraphics[width=8.5cm]{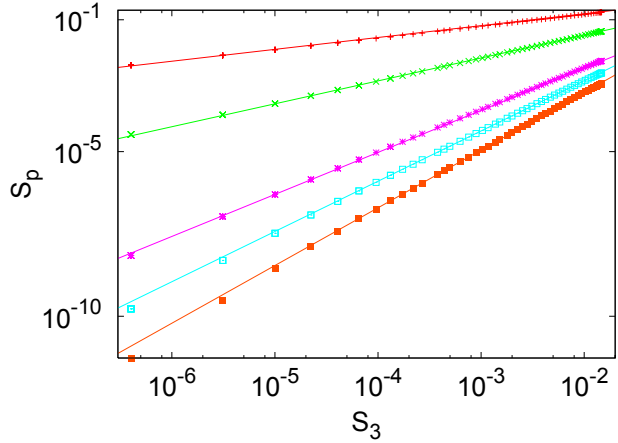}
  \caption{Eulerian structure functions of order p=1,2,4,5,6 (top to bottom)}
  \label{fig1}
\end{figure}

\begin{figure}
  \centering
  \includegraphics[width=8cm]{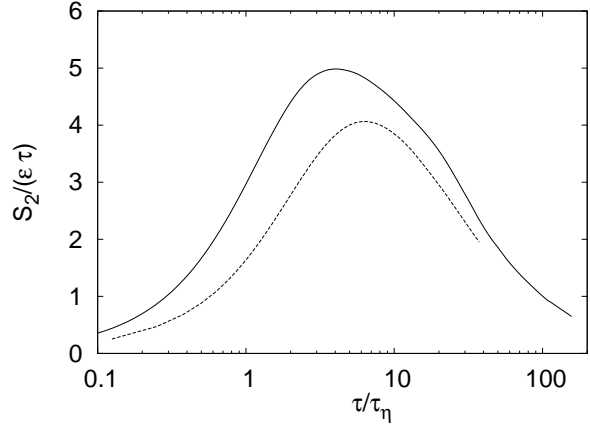}
  \caption{2nd order structure function normalized to $\epsilon \tau$
  for Run1 (solid line) and Run4 (dashed line)}
  \label{fig2}
\end{figure}

\begin{figure}
\centering
\includegraphics[width=9cm]{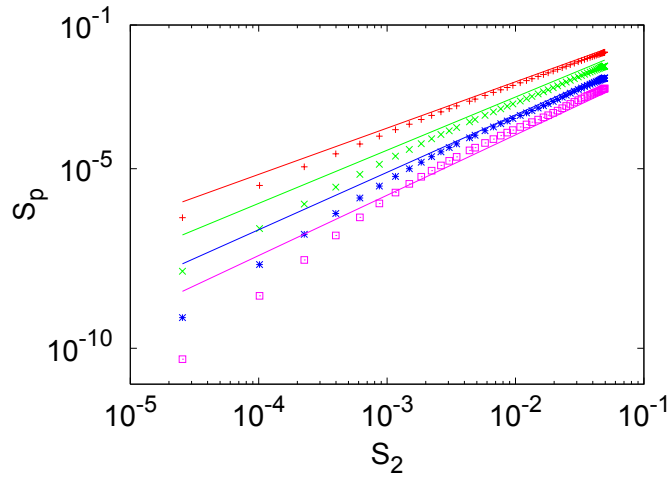}
\caption{Lagrangian structure functions from Run3 of order p=3-6 (top to bottom)}
\label{fig3}
\end{figure}

The exponents of the longitudinal Eulerian structure functions $S_p =
<|\mathbf{u}(\mathbf{x}+\mathbf{l})-\mathbf{u}(\mathbf{x})|^p>$,
angular brackets denoting spatial averaging, can be described by the
She-L\'ev\^eque formula~\citep{she-leveque:1994,dubrulle:1994} for the
simulations Run1 - Run3. The structure functions using extended self
similarity (ESS) \citep{benzi-ciliberto-etal:1993} together with
straight lines illustrating the fitted slopes are shown in
Figure~\ref{fig1} for data obtained from Run3. The values of the
corresponding exponents are summarized in Table~\ref{table2}. These
Eulerian statistics serve just as a test to check the numerics.
\begin{table}
  \centering
  \begin{tabular}{ccc}
    order &She-L\'ev\^eque &Run3 \\ \hline
    1     &0.364       &0.36 $\pm$ 0.0027 \\
    2     &0.696       &0.696 $\pm$ 0.0027 \\
    3     &1           &1 \\
    4     &1.279       &1.276 $\pm$ 0.0053 \\
    5     &1.538       &1.526 $\pm$ 0.013 \\
    6     &1.778       &1.752 $\pm$ 0.024 \\
    7     &2.001       &2.028 $\pm$ 0.088 \\
    8     &2.211       &2.204 $\pm$ 0.087      
  \end{tabular}
  \caption{Eulerian structure functions obtained using ESS with $\zeta_3 = 1$.}
  \label{table2}
\end{table}
The determination of the Lagrangian structure functions $S_p =
<|\mathbf{u}(t+\tau)-\mathbf{u}(t)|^p>$, angular brackets denoting
temporal averaging, turned out to be much more
difficult. Figure~\ref{fig2} shows a typical plot of the second order
Lagrangian structure function normalized to $\epsilon \tau$. It is
clear that no scaling range is present as already observed in the
experiments \citep{mordant-leveque-etal:2004,xu-bourgain-etal:2005} and
simulation \citep{biferale-etal:2004,yeung-borgas:2004}. Therefore, in
order to obtain scaling exponents one has to rely on the assumption of
extended self similarity. Figure~\ref{fig3} shows an evaluation of the
Lagrangian structure functions assuming ESS.  In Table~\ref{table3} we
present the relative exponents $\zeta_p/\zeta_2$ for our simulations.
In addition, this table contains also exponents collected from the
experiments \citep{mordant-leveque-etal:2004,xu-bourgain-etal:2005} and
other numerical simulations
\citep{biferale-etal:2004,mordant-leveque-etal:2004}.

\begin{table}
\centering
\begin{tabular}{l|l|l|l|l|l}
  order                                       &              1 & 3              & 4              & 5              & 6\\ \hline
  Run1                                        &0.58$\pm$0.006&1.28$\pm$0.020&1.46$\pm$0.06 &1.58$\pm$0.12 &1.67$\pm$0.19 \\
  Run2                                        &0.57$\pm$0.007&1.29$\pm$0.025&1.48$\pm$0.066&1.60$\pm$0.12 &1.68$\pm$0.18 \\
  Run3                                        &0.57$\pm$0.005&1.30$\pm$0.016&1.51$\pm$0.041&1.65$\pm$0.075&1.76$\pm$0.11 \\
  MF-NS                                       &0.55          &1.38          &1.71          &2.00          &2.26          \\
  Simulation~\citep{biferale-etal:2004}       & -            &-             & 1.7$\pm$0.05 &2.0$\pm$0.05  & 2.2$\pm$0.07  \\
  Experiment~\citep{mordant-leveque-etal:2004}& 0.56$\pm$0.01&1.34$\pm$0.02 &1.58$\pm$0.06 &1.76$\pm$0.1  & 1.9$\pm$0.2  \\
  Simulation~\citep{mordant-leveque-etal:2004}& 0.56$\pm$0.02&1.3 $\pm$0.04 & 1.5$\pm$0.09 &1.61$\pm$0.13 &1.69$\pm$0.2  \\
  Experiment~\citep{xu-bourgain-etal:2005}    & 0.58$\pm$0.12&1.28$\pm$0.30 &1.47$\pm$0.38 &1.59$\pm$0.46 &1.66$\pm$0.53   
\end{tabular}
\caption{Relative ESS-exponents calculated with respect to the structure function of order 2.
MF-NS denotes the Multi-Fractal approach for Navier-Stokes. Reynolds numbers for the different simulations and experiments:
Run1:                                         $R_\lambda = 190$,
Run2:                                         $R_\lambda = 122$,
Run3:                                         $R_\lambda = 178$,
Simulation~\citep{biferale-etal:2004}:        $R_\lambda = 284$,
Experiment~\citep{mordant-leveque-etal:2004}: $R_\lambda = 1000$,
Simulation~\citep{mordant-leveque-etal:2004}: $R_\lambda = 140$,
Experiment~\citep{xu-bourgain-etal:2005}:     $R_\lambda = 815$}

\label{table3}
\end{table}

First, we observe that the simulations Run1 - Run3 all give exponents
which agree within the error bars. Thus, a possible dependence of the
exponents on the Reynolds number or the choice of a stricter criterion
for the numerical resolution could not be detected given the relative
large error bars. In addition, the exponents fit quite well to the
present experiments
\citep{mordant-leveque-etal:2004,xu-bourgain-etal:2005} but are clearly
different from the exponents obtained by \citep{biferale-etal:2004}.
The reason for this is not the different interpolation for particle
velocities (trilinear in \citep{biferale-etal:2004}, tricubic here). We
repeated a simulation with the parameters of Run2 but using trilinear
interpolation and got the same scaling as with tricubic interpolation.
We explain this discrepancy similar as in \citep{xu-bourgain-etal:2005}
by observing that the evaluation of the structure functions was
performed systematically using larger values of $\tau$ compared to
us. We have have chosen the inertial range from Figure~\ref{fig2} by
the requirement that the function stays above ninety percent of its
maximum value. This leads to an inertial range of $2 \le \tau \le
7$. We are able to reproduce the exponents of \citep{biferale-etal:2004}
if we choose a minimal value of $\tau$ of about eight. Larger values
of $\tau$ result in a more Gaussian behavior and explains why the
exponents of \citep{biferale-etal:2004} are less intermittent.

\begin{figure}[!h]
\centering
\includegraphics[width=8cm]{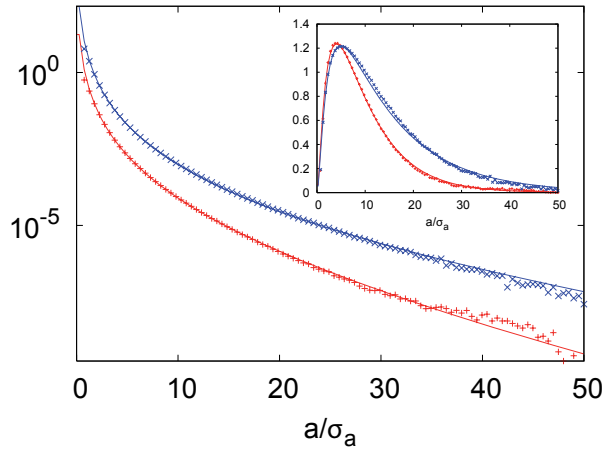}
\caption{Acceleration PDFs for runs Run2(+) and Run3(x) including the
  PDFs using the multifractal approach for the Navier-Stokes
  simulations. The inset shows the PDF multiplied by $(a/\sigma)^4$.}
\label{fig4}
\end{figure}

We also applied the multifractal model to the acceleration statistics
obtained from the simulations. We shortly review the approach of
\citep{biferale-etal:2004}. One starts with a suitable description for
the Eulerian structure functions, e.g. the She-L\'ev\^eque model
\citep{she-leveque:1994} and performs a Legendre transformation to
obtain the singularity spectrum. In order to translate Eulerian to
Lagrangian increments, one assumes a Kolmogorov like relation
$\delta_\tau v \sim \delta_l u$ where temporal and spatial increments
are related by $\tau_l \sim l/\delta_lu$. The resulting expression for
the Lagrangian structure functions,
\begin{equation}
  S_p(\tau) \sim \langle v_0^p \rangle \int_{h \in I} 
  dh \; \left( \frac{\tau}{T_L} \right)^{\frac{h p +3 -D(h)}{1-h}} \; ,
\label{lagrangesfunction}
\end{equation}
is evaluated by a saddle point
integration. To obtain the acceleration PDF, first the acceleration is
defined as $a = \frac{\delta_{\tau_\eta} v }{\tau_\eta}$ where
$\tau_\eta = \tau_\eta(h,u_0)$ is the Kolmogorov time scale which is
itself a multifractal quantity depending on the large scale velocity
field $u_0$. Assuming a Gaussian statistics of $u_0$ and integrating
over the possible scaling factors $h$ results in an explicit
expression for the acceleration PDF (see \citep{biferale-etal:2004} for
details)
\begin{eqnarray*}
  P(a) \sim \int_{h_{\mbox{\scriptsize min}}}^{h_{\mbox{\scriptsize max}}} dh \;
  &&\tilde{a}^{((h-5+D(h))/3)} R_\lambda^{y(h)} \\ && \exp \left( -
  \frac{1}{2} \tilde{a}^{2 (1+h)/3} R_\lambda^{z(h)} \right)
\end{eqnarray*}
with $\tilde{a} = a/\sigma_a$, $\sigma_a = \left< a^2 \right>^{1/2}$,
\mbox{$y(h) = \chi (h-5+D(h))/6 + 2(2D(h) +2h-7)/3$} and $z(h) = \chi
(1+h)/3 + 4(2h-1)/3$.
A comparison of the multifractal prediction to the numerically
obtained acceleration PDFs is shown in Figure~\ref{fig4}. Although
there is an excellent agreement between prediction and simulation, one
has to keep in mind that the multifractal prediction contains three
parameters.
The first parameter is hidden in the relation $\left< a^2 \right>
\propto R_\lambda^\chi$ when normalizing the width of the PDF.
However, this parameter can be determined from one simulation and
should then be kept fixed for other Reynolds numbers.
In the She-L\'ev\^eque model, the value of $h_{\mbox{\scriptsize min}}$ is given
by $h_{\mbox{\scriptsize min}}=1/9$. If one uses this value, it is not possible to
get good agreement with the measured shape of the PDF. Therefore, as
in \citep{biferale-etal:2004}, we use $h_{\mbox{\scriptsize min}}$ as a free
parameter. The last is a free amplitude in the normalization. In order
to get such an excellent agreement in Figure~\ref{fig4}, we had to
choose $h_{\mbox{\scriptsize min}} = 0.175$ for Run2 and $h_{\mbox{\scriptsize min}} = 0.16$
for Run3. The dependence on $h_{\mbox{\scriptsize max}}$ is negligible. \\

\noindent
\section{MHD turbulence}
The parameters for the MHD simulations are summarized in
Table~\ref{table1} (Run4, Run5). Both runs were performed with
negligible magnetic and cross helicity. 
\begin{figure}[!h]
\centering
\includegraphics[width=8.5cm]{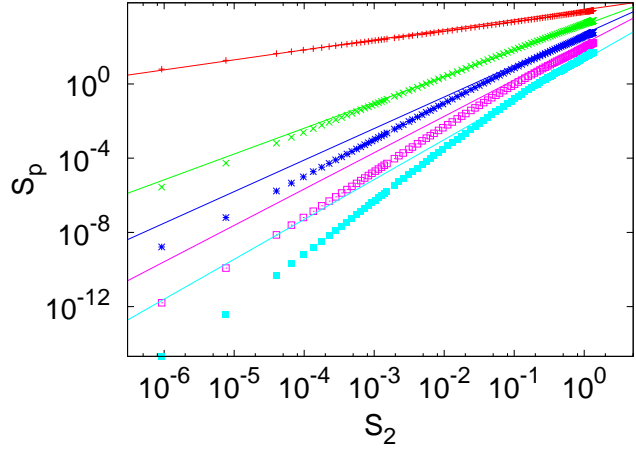}
\caption{Lagrangian structure functions from Run4 of order 1,3,4,5,6 (top to bottom)}
\label{fig5}
\end{figure}
1ESS plots of the Lagrangian velocity structure functions are shown in
Figure~\ref{fig5}. They show a similar curved shape although no
trapping in vortex tubes appears in MHD turbulence (see also Biferale
et al. \citep{biferale-etal:2005}).

The exponents for the Lagrangian velocity structure functions are
given in Table~\ref{table4}. Also shown is the prediction by a
multifractal model, which was obtained using the same steps as
described above, but starting with a She-L\'ev\^eque like formula
suitable for incompressible MHD-turbulence
\citep{horbury-balogh:1997,mueller-biskamp:2000},
\begin{equation}
  \zeta_L^{\mbox{\scriptsize MHD}}(p) = \frac{p}{9} + 1 - \left(\frac{1}{3}\right)^{p/3} \; .
  \label{msl}
\end{equation}
Although this formula is strictly valid only for the structure
functions of the Els\"asser variables $\mathbf{z}^\pm = \mathbf{u} \pm
\mathbf{B}$, we assume a cascade in the kinetic energy so that this
formula can also be applied to the structure functions of velocity.
The resulting multifractal model shows now an increased degree of
intermittency compared to the numerical simulations (Run4, Run5). On
the first sight this is astonishing since this is just the
opposite behavior as in the Navier-Stokes case. To summarize, we have
the following situation that in the Eulerian description, MHD
turbulence is more intermittent than Navier-Stokes turbulence whereas
the situation is reversed in the Lagrangian picture. This finding is
also not compatible with the multifractal ansatz. The multifractal
ansatz possesses a certain \textit{monotonicity} property. This means that if
for two different sets of structure function exponents, one is more
intermittent than the other in the Eulerian picture, than this one is
also more intermittent in the Lagrangian turbulence. To see this, it
is sufficient to look at high values $p$ of the order of the structure
functions. One observes that the value of $h^*$ where the infimum of
\begin{displaymath}
  h p + 3 - D(h)
\end{displaymath}
is assumed goes to $h_{\mbox{\scriptsize min}}$ for high values of $p$.  
Thus the asymptotic behavior reads
\begin{displaymath}
  \zeta_p = h_{\mbox{\scriptsize min}} p + 3 - D(h_{\mbox{\scriptsize min}}) \;\;\; , \;\; p \gg 1 \; .
\end{displaymath}
For the saddle point evaluation of the Lagrangian structure functions
(see eqn.  \ref{lagrangesfunction}) one has to find the infimum of
\begin{displaymath}
  \frac{h p + 3 - D(h)}{1-h} \;
\end{displaymath}
so that the asymptotic behavior is given by
\begin{displaymath}
  \zeta_p = \frac{h_{\mbox{\scriptsize min}} p + 3 - D(h_{\mbox{\scriptsize min}})}{1-h_{\mbox{\scriptsize min}}} 
  \;\;\; , \;\; p \gg 1 \; .
\end{displaymath}
Since both in Navier-Stokes and MHD the value of $h_{\mbox{\scriptsize min}}=1/9$
is identical, the degree of intermittency is determined by
$D(h_{\mbox{\scriptsize min}})$. This is both valid for the Eulerian as well as for
the Lagrangian model which guarantees the \textit{monotonicity} property.
\begin{table}
\centering
\begin{tabular}{c|l|l|l}
  order      &Run4              &Run5              &MF-MHD \\
  $R_\lambda$&187               &270               &       \\ \hline
  1          &0.527 $\pm$ 0.004 &0.526 $\pm$ 0.002 &0.63   \\
  2          &1                 &1                 &1      \\
  3          &1.412 $\pm$ 0.013 &1.407 $\pm$ 0.014 &1.26   \\
  4          &1.76 $\pm$ 0.04   &1.73 $\pm$ 0.06   &1.47   \\
  5          &2.06 $\pm$ 0.08   &1.96 $\pm$ 0.14   &1.65   \\
  6          &2.24 $\pm$ 0.24   &2.11 $\pm$ 0.25   &1.81
\end{tabular}
\caption{Relative ESS-exponents calculated with respect to the 
  structure function of order 2.}
\label{table4}
\end{table}
\begin{figure}[!h]
\centering
\includegraphics[width=4.8cm]{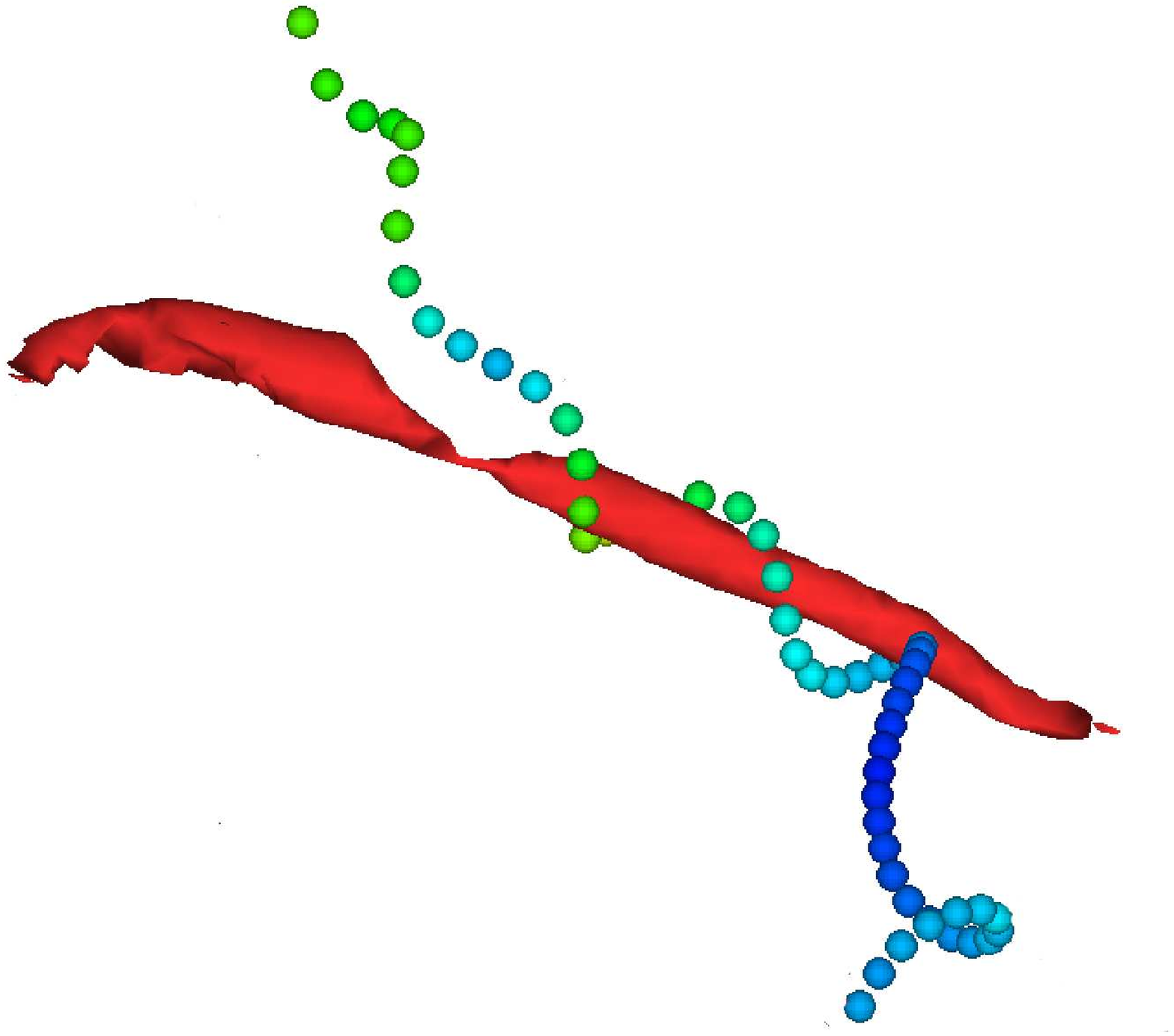} \hspace*{-0.5cm}
\includegraphics[width=3.8cm]{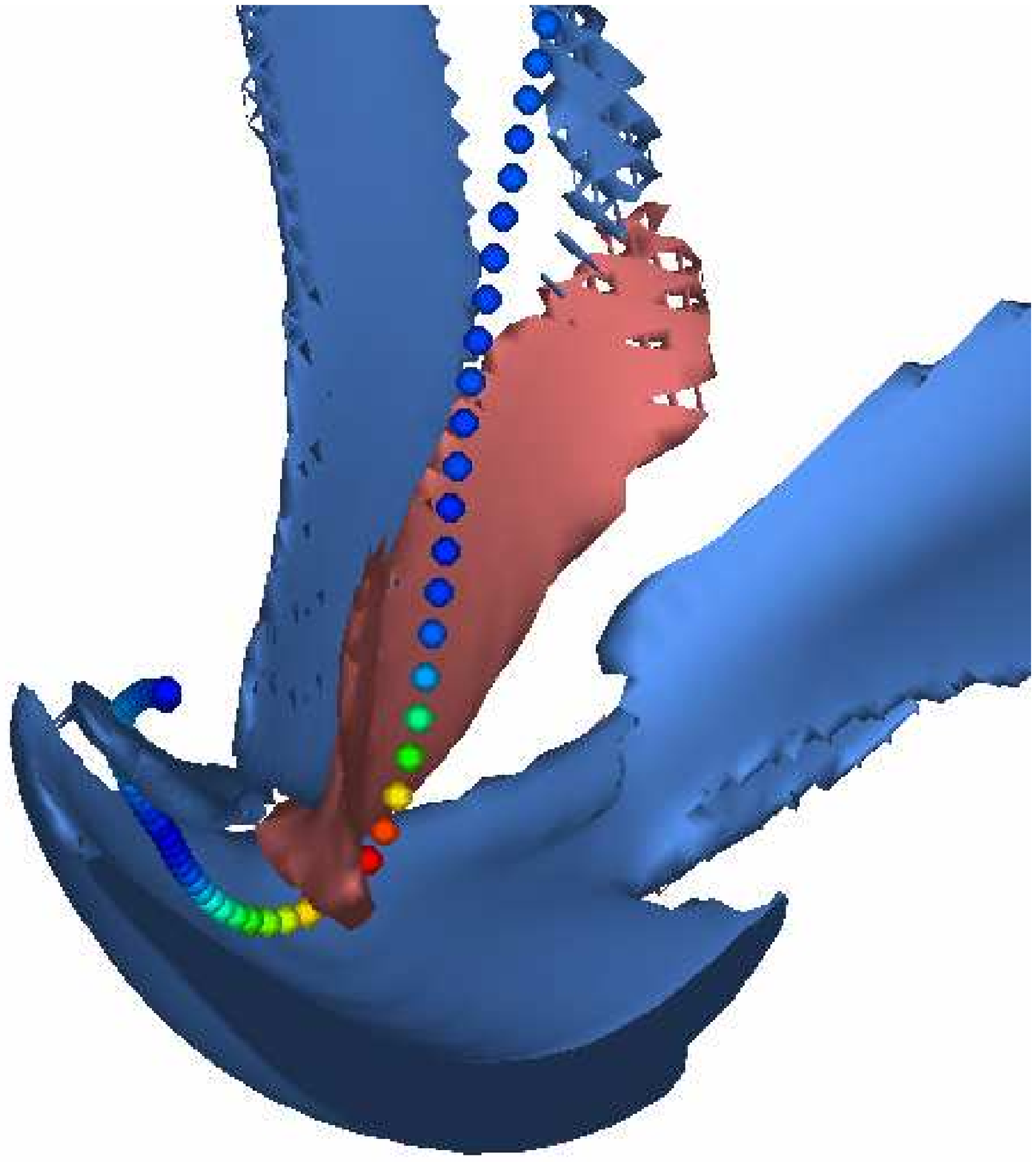}
\caption{Trajectories with high acceleration in NS (left) and MHD
  (right, blue denotes current density and red vorticity). }
\label{fig6}
\end{figure}

Our conclusion from this numerical observation is that the geometry of
the most singular structures (vortex tubes and current sheets) is not
the right quantity to determine the degree of Lagrangian
intermittency, but it is more important to look at the tracer
dynamics.  In Figure \ref{fig6} particle trajectories with high
acceleration near singular structures are shown. Here the isosurfaces
belong to a fixed point in time. Important is that in the MHD case the
trajectories near the sheet structures are smooth. Thus contrary to
the Eulerian point of view where the sheets are responsible for
producing intermittency, they do not contribute significantly to
Lagrangian intermittency. Large changes with high acceleration occur
at the ends of the sheet structures. Thus a naive translation of
Eulerian to Lagrangian structures is not possible. A more detailed
investigation of the relation between Lagrangian intermittency and the
small-scale structure of dissipation will be presented elsewhere.
\begin{figure}[!h]
\centering
\includegraphics[width=8.5cm]{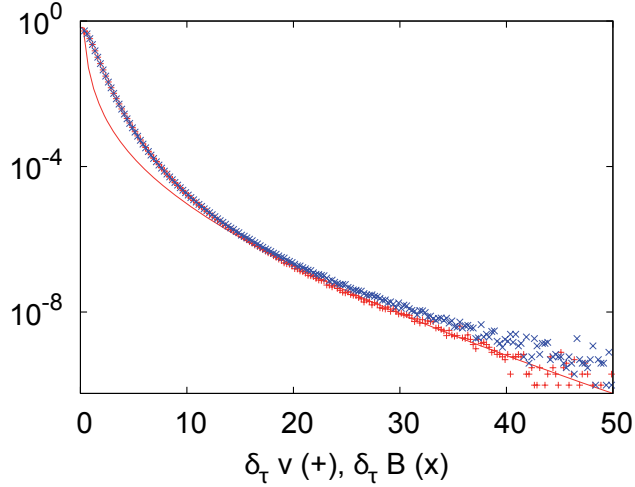}
\caption{Comparison of PDFs of velocity(+) and magnetic field(x)
  increments for small $\tau$: magnetic field increment are
  broader. The continuous line corresponds to the multifractal
  approach.}
\label{fig7}
\end{figure}
Using the multifractal approach we also compared the PDFs of velocity
and magnetic field increments of the order of the Kolmogorov time with
the multifractal prediction which is depicted in Figure~\ref{fig7},
again assuming the validity of (\ref{msl}) for the velocity and magnetic
field. Here, we have chosen $h_{\mbox{\scriptsize min}} = 0.16$ to obtain the
best agreement between the model and prediction. The agreement is not
as perfect as in the Navier-Stokes case. Here, only the exponential
tail could be well described by the multifractal model. \\

\noindent
\section{Conclusions and open questions} 
\begin{figure}[!h]
\centering
\includegraphics[width=8.5cm]{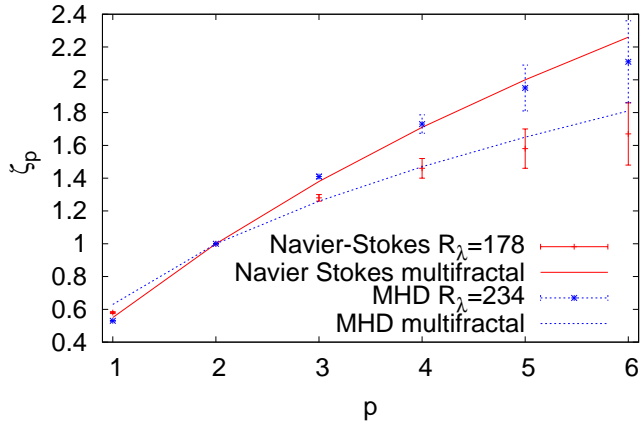}
\caption{Comparison of measured scaling exponents to the multifractal prediction}
\label{fig8}
\end{figure}

The presented Navier-Stokes
simulations show good agreement with recent experiments but deviate
from predictions of a multifractal model. An observation which also
could not be described by multifractal modeling is that Lagrangian
Navier-Stokes intermittency is stronger than in the MHD case whereas
the situation is reversed for Eulerian statistics. The present
situation is depicted in Figure~\ref{fig8}. It shows that the
multifractal prediction for Lagrangian Navier-Stokes turbulence fits
well to the simulations of Lagrangian MHD turbulence and vice versa.
This again shows that it is not easily possible to relate the geometry
of the most dissipative structures to the strong acceleration events.

A second and related critical issue in the Lagrangian treatment is the
validity of the Kolmogorov like relation $l \sim \tau u_l$ which
connects Eulerian and Lagrangian quantities. At least in the
neighborhood of strongly dissipating structures (tubes and sheets)
this relation has to be altered to $l \sim \tau u_0$ where $u_0$ is
the mean flow produced by the vortex. A similar reasoning
was given in \citep{li-meneveau:2005} to calculate the basic mechanism
for obtaining exponential tails in the PDF of velocity increments.

The partial success of the PDF modeling with the multifractal ansatz
is mainly attributed to the freedom of choosing three free parameters:
i) one parameter hidden in the normalization of the Reynolds number,
ii) the value of $h_{\mbox{\scriptsize min}}$ and finally iii) the
amplitude. If one fixes $h_{\mbox{\scriptsize min}}$ to the
She-L\'ev\^eque value $h_{\mbox{\scriptsize min}} = 1/9$ then this
modeling is not able to reproduce the shape of the PDF.

Thus a deeper understanding of the connection between the geometry of
dissipative structures and high acceleration events is necessary to
correctly model Lagrangian intermittency. \\

\noindent
{\it Acknowledgments.---}
Access to the JUMP multiprocessor computer at the FZ J\"ulich was made
available through project HB022. Part of the computations were
performed on an Linux-Opteron cluster supported by HBFG-108-291. A.B.
and WC.M. would like to thank the staff at RZG, Garching and LRZ,
Munich for competent and friendly assistance. The work of H.H. and
R.G. benefitted from support through SFB 591 of the Deutsche
Forschungsgesellschaft.

\end{document}